\documentclass[preprint,12pt]{revtex4-1}

\usepackage[brazil, english]{babel}
\usepackage[utf8]{inputenc}
\usepackage{graphicx}
\usepackage{epsfig}
\usepackage{amssymb}
\usepackage{amsmath} 
\usepackage{soul}
\usepackage[unicode=true,bookmarks=false,breaklinks=false,pdfborder={0 0 1},colorlinks=true]
 {hyperref}
\hypersetup{
 citecolor=blue,linkcolor=blue,urlcolor=blue}
\usepackage[mathscr]{euscript}
\usepackage{mathrsfs}
\usepackage{array}
\usepackage{slashed}
\usepackage{dsfont}
\usepackage[normalem]{ulem}
\usepackage{verbatim}

\begin{document}

\title{Fermionic DIS from a deformed string/gauge \\correspondence model}
\author{Eduardo Folco Capossoli$^1$}
\email{eduardo\_capossoli@cp2.g12.br}
\author{Miguel Angel Martín Contreras$^2$}
\email{miguelangel.martin@uv.cl}
\author{Danning Li$^{3}$}
\email{lidanning@jnu.edu.cn} 
\author{Alfredo Vega$^2$}
\email{alfredo.vega@uv.cl} 
\author{Henrique Boschi-Filho$^{4}$}
\email{boschi@if.ufrj.br}  
\affiliation{$^1$Departamento de F\'\i sica / Mestrado Profissional em Práticas de Educação Básica (MPPEB), Col\'egio Pedro II, 20.921-903 - Rio de Janeiro-RJ - Brazil\\
$^2$Instituto de Física y Astronomía, Universidad de Valpara\'iso, A. Gran Breta\~na 1111, Valpara\'iso, Chile\\
$^3$Department of Physics and Siyuan Laboratory, Jinan University, Guangzhou 510632, China\\
 $^4$Instituto de F\'\i sica, Universidade Federal do Rio de Janeiro, 21.941-972 - Rio de Janeiro-RJ - Brazil}

\begin{abstract}
From a deformed AdS$_5$ space, we used the string/gauge duality to study the deep inelastic scattering for unpolarized fermions with spin 1/2, considering the large Bjorken $x$ parameter regime. Here, we also took into account an anomalous dimension of an operator which represents fermions at the boundary. From this analysis, we compute the corresponding structure functions, which are dependent on $x$ and on the photon virtuality $q$. The results achieved are in agreement with the experimental data.
\end{abstract}


\maketitle

\section{Introduction}

Deep Inelastic Scattering (DIS) is an important tool in order to promote the fully understanding of the hadrons structure and its composition. As DIS belongs to the realm of strong interactions the appropriate theory to deal with it is the $SU(3)$ gauge field theory called as QCD.  From the beginning of the QCD itself some important works, for instance in Refs. \cite{Georgi:1951sr, Gross:1976xt}, relate that the insertion of an anomalous contribution $\gamma$ can change the canonical dimension $\Delta_{\rm can}$ of an operator $\cal O$, such as, $[{\cal O}]=\Delta_{\rm can} + \gamma$ and relate this anomalous dimension with the  DIS. 

In this work, we will explore the fact the  perturbative techniques applied to QCD do not work well at low energy regimes, and then, we will study DIS from an alternative perspective known as anti-de Sitter/Conformal Field Theory (AdS/CFT) correspondence. This correspondence relates superstring theory in a 10-dimensional curved spacetime to a conformal ${\cal N} = 4$ super Yang-Mills theory (SYM) with symmetry group $SU(N)$, for $N \rightarrow \infty$, living in $3+1$ dimensional Minkowski spacetime. For an important  review, one can see Ref. \cite{Aharony:1999ti}. 

This correspondence becomes fully applicable after breaking the conformal symmetry. By doing this, one can build a phenomenological holographic approach known as AdS/QCD. In this field, one has many important works dealing with DIS and bringing new understandings and results. This can be seen for instance in Refs. \cite{Polchinski:2002jw, BallonBayona:2007rs, BallonBayona:2007qr, Gao:2009ze, Gao:2010qk, BallonBayona:2010ae, Braga:2011wa, Gao:2014nwa, Capossoli:2015sfa}. This set of reference includes the studies of DIS  for scalar, mesonic and baryonic particles. The important Ref. \cite{Polchinski:2002jw} dealt with the holographic DIS within hardwall model for scalars and fermions considering large, small and exponentially small Bjorken parameter $x$ scenarios. Vector particle DIS was studied, for instance in Ref. \cite{BallonBayona:2010ae}, whilst in Refs. \cite{BallonBayona:2007qr, Gao:2009ze, Gao:2010qk, Braga:2011wa} studied DIS for baryons.

Here, throughout AdS/CFT duality, using a deformed background holographic model we will calculate the proton structure functions, in the large $x$ regime, $F_1 = F_1 (x, q^2)$ and  $F_2 = F_2 (x, q^2)$. These structure functions are written as a function of the photon virtuality $q^2$, and the Bjorken parameter $x$. 

In particular, our AdS/QCD model with a deformed $AdS_5$ background  breaks the conformal symmetry producing a mass gap for the fermionic fields. Deformed holographic models were used for several purposes within AdS/QCD context, as can be seen in Refs. \cite{Andreev:2006ct,  Afonin:2012jn, Rinaldi:2017wdn, Bruni:2018dqm, Afonin:2018era,  FolcoCapossoli:2019imm, MartinContreras:2021yfz,  Contreras:2021epz}.

This work is organized as follows: in section \ref{secdis} we describe the DIS phenomenology.  In section \ref{deformed} we present our deformed AdS space model where we compute the electromagnetic and the fermionic fields.  In section \ref{disintaction} we calculate the DIS interaction action and derive the corresponding structure functions $F_1(q^2, x)$ and $F_2(q^2, x)$. In Section \ref{numerical} we present our numerical results for the structure functions and compare them with available experimental data. In Section \ref{conc} we present our conclusion.

\section{Deep Inelastic Scattering  phenomenology} \label{secdis}
This kind of scattering processes occurs in high-energy regimes and it is achieved by colliding a lepton ($\ell$) and a proton ($p$), through the exchange of a virtual photon, which is represented by $\ell p \to \ell X$. The many fragments produced are represented by $X$. 

In the context of the scattering theory, DIS is ruled by some parameters, such as the photon virtuality $q$ and the Bjorken variable $x$, given by:
\begin{equation}
    x= - \frac{q^2}{2\, P \cdot q}\,,
\end{equation}
where $P$ is the initial proton momentum, defined as $P^2 = -M^2$.

The DIS scattering amplitude can be written as:

\begin{equation}
i\,\mathcal{M}_{l\,p\to l\,X} \propto  \int{d^4y\,e^{i\,q\cdot y}\langle X\left|J^\mu (y)\right|P\rangle},
\end{equation}
\noindent where the electromagnetic current associated to quark is represented by $J_\mu(x)$.

By using the optical theorem one can write the transition amplitude $W^{\mu\,\nu}$ or  \emph{the forward Compton scattering amplitude} as a function of the  forward matrix element of two proton currents averaged over the spin, so that:
\begin{equation}\label{f3}
W^{\mu \nu} = \frac{i}{4\, \pi} \sum_s \int d^4 y~ e^{i q.y} \langle P, s|\mathcal{T}\left\{J^{\mu}(y)\, J^{\nu}(0)\right\}|P, s \rangle\,,
\end{equation}
\noindent where $|P, s \rangle$ are the normalizable proton states, considering it with spin $s$. For a review one can see Ref. \cite{Manohar:1992tz}. 

By taking the imaginary part of the transition amplitude, written in a Fourier space, and taking into account the following condition $q_\mu\,W^{\mu\nu}=q_\nu\,W^{\mu\nu}=0$, the hadronic tensor has a decomposition given by:
\begin{eqnarray}\label{f4}
W^{\mu \nu} &=& F_{1} \left( \eta^{\mu \nu} - \frac{q^{\mu} q^{\nu}}{q^2} \right)\nonumber \\ &+& \frac{2x}{q^2} F_2 \left( P^{\mu} + \frac{q^{\mu}}{2x} \right)\left( P^{\nu} + \frac{q^{\nu}}{2x} \right).
\end{eqnarray}

The scalar functions $F_{1,2}\equiv F_{1,2}(x,q^2)$ are the DIS Structure Functions that we are going to compute holographically. Besides for our purpose, we are going to consider only unpolarized leptons and target protons.

\section{The deformed string/ gauge correspondence model and the DIS}\label{deformed}

The deformed string/gauge model, used in this work, can be considered as a variation of the original AdS/CFT correspondence. In the deformed version, we have introduced an exponential factor $e^{k z^2}$ in the AdS$_5$ metric. This deformation suggests that we should take into account an anomalous dimension in this approach.

In our model, the bulk field takes the following form:
\begin{equation}\label{acao_soft}
S = \int d^{5} x \sqrt{-g} \; {\cal L}\,,
\end{equation}

\noindent where ${\cal L}$ is the Lagrangean density, and $g$ is the determinant of the metric $g_{mn}$ of the deformed $AdS_5$ space, given by:
\begin{equation}\label{gs}
ds^2 = g_{mn} dx^m dx^n= e^{2A(z)} \, (dz^2 + \eta_{\mu \nu}dy^\mu dy^\nu)\,. 
\end{equation}

\noindent Note that we set the AdS radius $R=1$.
\begin{equation}
    A(z) = -\log z + \frac{k}{2}\, z^2\,, \label{A}
\end{equation}
\noindent with $z$ the holographic coordinate and where the constant $k$ has dimension of GeV$^2$ which is associated with a QCD mass scale. Here, we take  $
m, n, \cdots$  to refer to the AdS$_5$ space. We also separate into $\mu, \nu, \cdots$ for the Minkowski spacetime and the holographic $z$ coordinate. The metric $\eta_{\mu \nu}$ has signature $(-,+,+,+)$ and will describe the gauge theory (boundary) of the deformed AdS space. 

It is worthwhile to mention that the deformed string/gauge model presented here is different from the original softwall model (SWM)  \cite{Karch:2006pv}. Here, in the deformed model the conformal factor is introduced in the AdS$_5$ metric instead in the action as in the SWM. As a consequence of this, the deformed model brakes fermion's conformal invariance producing discrete spectrum for fermions whilst SWM does not.

The deformed string/gauge correspondence model is based on Refs. \cite{Andreev:2006ct}. The same deformation in the AdS$_5$ was applied to compute the spectrum of several particles, with various spins, also including spin 1/2 fermions \cite{FolcoCapossoli:2019imm}. This fermionic computation is closely related to the present discussion of DIS with fermionic target. 

The holographic description of DIS used in this work is inspired by Ref.  \cite{Polchinski:2002jw} considering large $x$ regime or in the supergravity approximation. From such a description, the AdS/CFT duality connects the matrix element of DIS presented in Eq.\eqref{f4} with the supergravity interaction action in AdS space, $S_{\rm int}$, so that:
\begin{equation}\label{corrint}
   \eta_\mu \langle P + q, s_X | J^{\mu}(0)|P, s\rangle = S_{\rm int},
\end{equation}
\noindent here, we are considering that the fermionic particle was scattered off by a virtual photon with polarization $\eta_{\mu}$. On the other hand, the interaction action reads:
\begin{eqnarray}\label{sintdis}
S_{\rm int}= g_V\,\int{dz\,d^4y\,\sqrt{-g}\,\phi^\mu\,\bar{\Psi}_X\,\Gamma_\mu\,\Psi_i}\,,
\end{eqnarray}
\noindent whith $g_V$ a coupling constant which depends on the electric charge of the fermion and $\Gamma_{\mu}$ are Dirac gamma matrices placed in curved space. $\phi^{\mu}$ is the massless electromagnetic gauge boson field and the spinors $\Psi_i$ and $\Psi_X$ are the initial and final states for the fermion.

\subsection{Calculating the electromagnetic field}

In this section we will present the massless gauge field in the deformed AdS space. It will represent the virtual photon at UV, responsible the electromagnetic interaction, needed in DIS.

The action for a five dimensional massless gauge field $\phi^n$ is written as:
\begin{equation}\label{f15}
S = -  \int d^{5} x \sqrt{-g} \; \frac{1}{4} F^{mn} F_{mn}\,,
\end{equation}
\noindent where $F^{mn} = \partial^m \phi^n - \partial^n \phi^m$ and the following equations of motion, given by $\partial_m [ \sqrt{-g}\; F^{mn}] = 0\,. $
\noindent By Considering the gauge fixing as follows:
\begin{equation}
\partial_\mu\,\phi^\mu+e^{-A}\partial_z\left(e^A\,\phi_z\right)=0,  
\end{equation}
where $A=A(z)$ is given by Eq. \eqref{A}, one gets after some algebra
\begin{eqnarray}\label{phimunorm}
\phi_\mu(z,q)&=& -\frac{\eta_{\mu} e^{i q\cdot y}}{2}\, B(z,q)\,, 
\end{eqnarray}
\noindent where $\Gamma[a]$ is the Gamma function and $B(z,q) \equiv {\cal U} \left(1-q^2/2 k;\,2;\,-kz^2/2\right)$ is the Tricomi hypergeometric function.

\subsection{Calculating the fermionic states} \label{barstates}

The action for the fermionic fields in the deformed AdS space is given by \cite{Henningson:1998cd}:  
\begin{equation}\label{diracfield}
S =  \int d^{5} x\sqrt{g} \; \bar{\Psi}({\slashed D} - m_5 ) \Psi, 
\end{equation}
From this action one can derive the corresponding equation of motion given by $({\slashed D} - m_5 ) \Psi=0$, where ${\slashed D}$ is defined as:
\begin{equation}\label{slash}
{\slashed D} \equiv   e^{-A(z)} \gamma^5 \partial_5 + e^{-A(z)} \gamma^{\mu}  \partial_{\mu} + 2 A'(z)\gamma^5, 
\end{equation}
\noindent where $\partial_5 \equiv \partial_z$, and $m_5$ is the baryon bulk mass. Now, one can decompose the spinor $\Psi$ into right- and left-handed chiral components, and applying some math, one has
\begin{eqnarray}\label{scr}
-\psi_{R/L}''(z) + \left[V_{R/L}(z) \right]\psi_{R/L}(z)  = M_n^2 \psi_{R/L}^n (z)\,.
\end{eqnarray}
\noindent Note that the above equation represents a Schr\"odinger equation which considers right and left sectors. Also note that $M_n$ is the boundary  baryon mass for each mode $\psi^n_{R/L}$ and the corresponding potentials is written as:
\begin{equation}\label{potscr}
 V_{R/L}(z) = m_5^2 e^{2 A(z)} \pm m_5 e^{A(z)}A'(z). 
\end{equation}
By using the solutions provided by Eq. \eqref{scr} one can read the final spinor state $\Psi_X$ and the initial spinor state  $\Psi_i$. These are linear combinations of the chiral solutions $\psi_{R/L}$. Besides $s_i$ and $s_X$ are the spin of the initial and final states, respectively.

Regarding the bulk mass, one can relate it to the conformal dimension $\Delta_{\rm can}$ of a boundary operator ${\cal O}$. In pure AdS$_5$ space, one has:
\begin{equation}
|m_5^{\rm AdS}|= \Delta_{\rm can} - 2. 
\end{equation}
Within the holographic context the concept of the anomalous dimension has appeared in some studies, for instance, Refs. \cite{Gubser:2008yx, Vega:2008te,  FolcoCapossoli:2016ejd, BoschiFilho:2012xr, Capossoli:2016ydo}.  Taking into account those cases, the effective conformal dimension is $\Delta_{\rm eff}=\Delta_{\rm can} + \gamma$, and then, on can write $|m_5|= \Delta_{\rm can} + \gamma - 2\,, $ where $\gamma$ is the anomalous contribution.

In this work we will resort to the anomalous dimension to calculate the structure functions in fermionic  DIS. Besides, we will follow \cite{pdg} where one can see that if we are dealing with the Bjorken variable $x \to 1$ (large $x$ regime) the parton distribution functions (PDFs) seem to indicate that the proton can be considered as a single particle (disregarding the internal constituents). This is applied in Ref. \cite{Braga:2011wa}, with $\Delta_{\rm can} = 3/2$. 

\section{The DIS interaction action}\label{disintaction}

Here, we will compute the DIS interaction action by using Eqs. \eqref{corrint} and \eqref{sintdis}, together with the final and the initial spinor states $\Psi_X$ and $\Psi_i$, respectively, so that:
\begin{eqnarray}\notag
S_\text{int}&=&\frac{g_V}{2}\,\left(2\,\pi\right)^4\,\delta^4(P_X-P-q)\,\,\eta^\mu\,\nonumber \\
&\times &\left[\bar{u}_{s_X}\,\gamma_\mu\,\hat{P}_R\,u_{s_i}\,\mathcal{I}_L+\bar{u}_{s_X}\,\gamma_\mu\,\hat{P}_L\,u_{s_i}\mathcal{I}_R\right]\label{sintfinal}
\end{eqnarray}

\noindent where the $\mathcal{I}_{R/L}$ are defined in terms of the solutions of the chiral fermions and the solution of the field $B$, so that:
\begin{equation}\
\mathcal{I}_{R/L}=\int{dz\,B(z,q)\,\psi_{R/L}^X(z,P_X)\,\psi_{R/L}^i(z,P)}\,. 
\end{equation}
As we interested in a spin independent scenario, and by summing over spin, one can write the structure functions, such as:
\begin{eqnarray}\label{F1}
F_1(q^2,x)&=&\frac{{g}^2_\text{eff}}{4}\,\Bigg[M_0\,\sqrt{M_0^2+q^2\left(\frac{1-x}{x}\right)}\,\mathcal{I}_L\,\mathcal{I}_R\nonumber \\&+&\left(\mathcal{I}_L^2+\mathcal{I}_R^2\right)\left(\frac{q^2}{4\,x}+\frac{M_0^2}{2}\right)\Bigg]\frac{1}{M_{X}^2}\\
F_2(q^2,x)&=&\frac{{g}^2_\text{eff}}{8}\frac{q^2}{x}\left(\mathcal{I}_L^2+\mathcal{I}_R^2\right)\frac{1}{M_{X}^2},\label{F2}
\end{eqnarray}
\noindent where $M_{X}\equiv M_{X}(q^2,x)$ is mass of the effective final hadron related to the mass of the initial hadron: 
\begin{equation}
    M_{X}^2(q^2,x) = M_0^2+q^2\,\left(\frac{1-x}{x}\right)\,. 
\end{equation}


\section{Numerical results for the structure functions}\label{numerical}

Here, we will summarize our numerical results related to the structure functions $F_1(x, q^2)$ and $F_2(x, q^2)$, and considering some values of the Bjorken parameter, such as, $x=0.65, 0.75, 0.85$. Besides, we compare our results with the available experimental data, as presented in Ref. \cite{Whitlow:1991uw}, from SLAC and in Ref. \cite{Benvenuti:1989rh}, from BCDMS.

In order to get our results, we solved numerically Eqs. \eqref{scr} and \eqref{F2}. In the figure below one can see the main results achieved in this work. This figure shows the structure function $F_2(x, q^2)$ as a function of  $q^2$ considering  $x=0.65$ with $m_5= 0.505$ GeV, $k = 0.374$ GeV$^2$, $g^2_{\rm eff} = 1.83$ and $\gamma$ = 0.005; $x=0.75$ with $m_5= 0.565$ GeV, $k = 0.340$ GeV$^2$, $g^2_{\rm eff} = 1.65$ and $\gamma$ = 0.065, and $x=0.85$ with $m_5= 0.878$ GeV, $k = 0.196$ GeV$^2$, $g^2_{\rm eff} = 1.83$ and $\gamma$ = 0.378. 

Note that these values of $m_5$,  $k$, $g^2_{eff}$ and $\gamma$ for each value of $x$ assure that the computed proton mass, from Eq. \eqref{scr}, is $m_p$= 0.938 GeV.

\begin{figure}
\centering
\includegraphics[scale = 0.65]{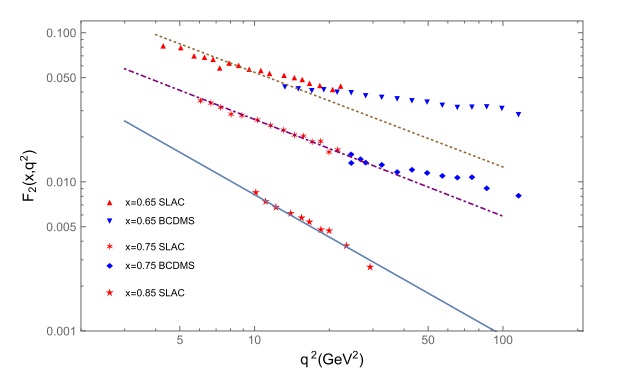} 
\caption{Here we are presenting our results for the structure function $F_2(x,q^2)$ in terms of $q^2$.  Also in this figure, we have considered the cases for $x = 0.65$,  $x = 0.75$ and $x = 0.85$ and its corresponding theoretical fitting represented by the dotted, dot-dashed, and solid lines, respectively. Notice that we also brought in this figure the experimental data from SLAC \cite{Whitlow:1991uw} and BCDMS \cite{Benvenuti:1989rh}.}
\label{results}
\end{figure}

\section{Conclusions} \label{conc}

In this section we will present our last comments on our holographic description for the DIS structure functions $F_1(x,q^2)$ and $F_2(x,q^2)$ for baryons. The present work is based on a deformed AdS metric and takes into account an anomalous contribution to the canonical conformal dimension of a CFT operator. 

The results for $F_2(x,q^2)$ when compared with ones coming from experimental data are in agreement. In particular, for large $x$ (or $x=0.85$) regime.

The complete calculations done in this work, including all details, can be seen in Ref. \cite{FolcoCapossoli:2020pks}.

\section*{Acknowledgments}

We would like to thanks Alfonso Ballon Bayona for useful discussions. 
E.F.C. and M.A.M.C would like to thank the hospitality of the Jinan University where part of this work was done. E.F.C also would like to thank the hospitality of the Valpara\'iso University where part of this work was done.   A. V. and  M. A. M. C.  would like to thank the financial support given by FONDECYT (Chile) under Grants No. 1180753  and No. 3180592,  respectively. D.L. is supported by the National Natural Science Foundation of China (11805084), the PhD Start-up Fund of Natural Science Foundation of Guangdong Province (2018030310457) and Guangdong Pearl River Talents Plan (2017GC010480). H.B.-F. is partially supported by Coordena\c c\~ao de Aperfei\c coamento de Pessoal de N\'ivel Superior (CAPES),  and Conselho Nacional de Desenvolvimento Cient\'ifico e Tecnol\'ogico (CNPq) under Grant No. 311079/2019-9.



\end{document}